\documentclass[%
 reprint,
superscriptaddress,
 amsmath,amssymb,
 aps,
]{revtex4-2}

\usepackage{graphicx}
\usepackage{dcolumn}
\usepackage{bm}
\usepackage{hyperref}
\usepackage[mathlines,modulo]{lineno}
\usepackage{booktabs}
\usepackage{sidecap}
\usepackage{hhline}

\usepackage{mfirstuc} 
\newcommand{\addReviewer}[2]{
  \expandafter\newcommand\csname #1\endcsname[1]{{\it{ \color{#2} \capitalisewords{#1}:\,##1}}}
  \expandafter\newcommand\csname #1cor\endcsname[2]{{\color{#2} \capitalisewords{#1}:\,\st{##1}{\it{##2}}}}
  \expandafter\newcommand\csname #1color\endcsname{#2}
  \expandafter\newcommand\csname #1todo\endcsname[1]{{\todo[inline,color=white!70!#2, caption={}]{\textbf{\capitalisewords{#1}}: ##1}}}
}

\usepackage{tikz, marginnote} 
\newcommand{\checkedby}[1]{
\ifdefined\CROSSCHECKS
  \marginnote{
    \begin{tikzpicture}
      \foreach \x [count=\xi] in {#1} {
         \node[shape=circle,inner sep=0mm,
         minimum size=2mm,
         fill=\csname \x color\endcsname] at (\xi*3mm,0) {};
       }
    \end{tikzpicture}
  }
\else
\fi
}

\usepackage{soul,color}
\definecolor{chromeyellow}{rgb}{1.0, 0.65, 0.0}
\definecolor{DodgeBlue}{rgb}{0.118, 0.565,1.000}
\definecolor{asparagus}{rgb}{0.53, 0.66, 0.42}
\definecolor{cadmiumgreen}{rgb}{0.0, 0.42, 0.24}
\definecolor{blue(ryb)}{rgb}{0.01, 0.28, 1.0}
\definecolor{periwinkle}{RGB}{181, 146, 203}
\definecolor{turquoiseblue}{rgb}{0.02, 0.55, 0.55}
\definecolor{atlanticstorm}{RGB}{57,107,127}
\definecolor{amethyst}{rgb}{0.6, 0.4, 0.8}
\definecolor{blue-violet}{rgb}{0.54, 0.17, 0.89}
\definecolor{brightlavender}{rgb}{0.75, 0.58, 0.89}
\definecolor{brightmaroon}{rgb}{0.76, 0.13, 0.28}
\definecolor{brilliantrose}{rgb}{1.0, 0.33, 0.64}
\definecolor{byzantine}{rgb}{0.74, 0.2, 0.64}
\definecolor{darkmagenta}{rgb}{0.55, 0.0, 0.55}

\addReviewer{johna}{blue}
\addReviewer{jq}{cyan}
\addReviewer{xx}{magenta}

\addReviewer{jianping}{red}
\addReviewer{xiaochao}{magenta}
\addReviewer{mike}{blue}

\addReviewer{aaa}{brown}
\addReviewer{bbb}{magenta}
\addReviewer{eee}{asparagus}
\addReviewer{fff}{periwinkle}
\addReviewer{hhh}{byzantine}
\addReviewer{iii}{brightmaroon}
\addReviewer{jjj}{pink}

\begin{document}

\title{Quasielastic $\overrightarrow{^{3}\mathrm{He}}(\overrightarrow{e},{e'})$ Asymmetry in the Threshold Region}

\author{M.~Nycz}
\email{mnycz@jlab.org}
\affiliation{University of Virginia, Charlottesville, Virginia 22904, USA} 
\author{W.~Armstrong}
\affiliation{Argonne National Laboratory, Lemont, Illinois 60439, USA}
\author{T.~Averett}
\affiliation{William \& Mary, Williamsburg, Virginia 23187, USA}
\author{C.Ayerbe Gayoso}
\affiliation{William \& Mary, Williamsburg, Virginia 23187, USA}
\author{X.~Bai}
\affiliation{University of Virginia, Charlottesville, Virginia 22904, USA}
\author{J.~Bane}
\affiliation{University of Tennessee, Knoxville, Tennessee 37996, USA}
\author{S.~Barcus}
\affiliation{Thomas Jefferson National Accelerator Facility, Newport News, Virginia 23606, USA}
\author{J.~Benesch}
\affiliation{Thomas Jefferson National Accelerator Facility, Newport News, Virginia 23606, USA}
\author{H.~Bhatt}
\affiliation{Mississippi State University, Mississipi State, Mississippi 39762, USA}
\author{D.~Bhetuwal}
\affiliation{Mississippi State University, Mississipi State, Mississippi 39762, USA}
% ...



\date{\today}

\begin{abstract}
A measurement of the double-spin asymmetry from electron-$^{3}$He scattering in the threshold region 
of two- and three-body breakup of $^{3}$He was performed at Jefferson Lab, for Q$^{2}$ values of 0.1 and 0.2 (GeV/$c$)$^{2}$. The results of this measurement serve as a stringent test of our understanding of few-body systems. When compared with calculations from plane wave impulse approximation and Faddeev theory, we found that 
the Faddeev calculations, which use modern nuclear potentials and prescriptions for meson-exchange currents, demonstrate an overall good agreement with data. 




\end{abstract}

\maketitle


\section*{\label{sec:level1}}

As the simplest nuclear environment that involves dynamics beyond the two-nucleon interactions, the $^3$He
\vfill\null
nucleus provides a testing ground of our understanding of the three-nucleon force (3NF), sometimes highlighting flaws or omissions~\cite{Miha_2014}. 
Significant theoretical effort has been made to accurately characterize three-nucleon systems, 
including two major methods that describe the process of electron scattering off A=3 nuclei: plane wave impulse approximation (PWIA)~\cite{PWIA1,PWIA2,PhysRevC.64.055203} and exact non-relativistic Faddeev calculations~\cite{Golak1995,GOLAK200589,Deltuva2004,DELTUVA2022137552,witała2023inclusion}. 
The PWIA typically has shown a good agreement with data at large four-momentum transfers. At low 
four-momentum transfers 
above the two-body ($p+d$) and three-body ($p+p+n$) breakup threshold,
non-relativistic Faddeev calculations that fully include Final State Interactions (FSI) and use realistic nucleon-nucleon potentials along with the Meson Exchange Currents (MEC), have shown a good agreement with data. 

Experimental measurements that can directly probe FSI, 3NF, and MEC, would provide additional limits and further refinements to theoretical models. 
This can be done by identifying kinematic regions where effects due to these three mechanisms are expected to be large. 
{Inclusive unpolarized cross section measurements made in the threshold region showed good agreement with theoretical predictions \cite{PhysRevC.49.1263}. 
On the other hand, initial measurements of spin observables 
suggested missing elements in theoretical description, possibly due to MEC and FSI, which was improved in later, more precise measurements~\cite{Feng_Xiong}. 

We report here a new, precision measurement of the double-spin asymmetry of electron scattering off the $^3$He in the elastic and quasielastic region in the threshold region. 
%
In such regions of low momentum transfer, rescattering of the nucleon which interacted with the virtual photon with the remaining nucleons can occur~\cite{Golak1995}, providing a unique testing ground of FSI and MEC.  
In the following, we describe the formalism, the experiment, data analysis, and the results.

For quasielastic scattering in which longitudinally polarized electrons scattering off a polarized $^{3}$He target, the inclusive differential cross section can be expressed as \cite{Donnelly_pol}: 
\begin{multline}    
    \frac{d\sigma}{d\Omega dE'} =\frac{\alpha^{2} \mathrm{cos}^{2}(\frac{\theta}{2})}{4 E^{2} \mathrm{sin}^{4}(\frac{\theta}{2})} \{ v_{l} R_{L} + v_{t} R_{T} \\
    - h \left[ v_{T^{'}}\mathrm{cos}\theta^{*} R_{T^{'}} + 2v_{TL^{'}} \mathrm{sin}\theta^{*} \mathrm{cos}\phi^{*} R_{TL^{'}}  \right] \},
    \label{eq:cross_section}
\end{multline}
where $\alpha$ is the fine-structure constant, $E$ is the initial energy, $\theta$ is the scattered electron angle,} $R_{T}$ and $R_{L}$ are the spin-averaged response functions, $R_{T^{'}}$ and $R_{TL^{'}}$ are the spin-dependent response functions, the terms $v_{i}$($i=L,T,T^{'},TL^{'}$) are kinematic factors, and $\theta^{*}$ and $\phi^{*}$ are the polar and azimuthal angles of the polarization vector of the nuclear target in the lab frame, with $\hat{z}$ pointing along the virtual photon momentum, $\hat{q}$~\cite{Elastic_Proposal}. 

The spin-dependent asymmetry is defined as 
\begin{equation}
    A = \frac{\sigma^{\uparrow\Uparrow} - \sigma^{\downarrow\Uparrow} }{ \sigma^{\uparrow\Uparrow} + \sigma^{\downarrow\Uparrow}},
    \label{eq:Asym_1}
\end{equation}
where 
$\sigma^{\uparrow\Uparrow}$ and $\sigma^{\downarrow\Uparrow}$ are the cross section for 
scattering with the electron spin parallel and anti-parallel to the target spin, 
respectively. Substituting the expression for the cross section given in Eq.(\ref{eq:cross_section}) into Eq.(\ref{eq:Asym_1}), the asymmetry can be written as:
\begin{equation}
    A = - \frac{\mathrm{cos}\theta^{*} v_{T^{'}} R_{T^{'}} + 2 \mathrm{sin}\theta^{*} \mathrm{cos}\phi^{*} v_{TL^{'}} R_{TL^{'}} }{v_{L}R_{L} + v_{T} R_{T}}.
    \label{eq:Asym_2}
\end{equation}


The experiment reported here~\cite{Elastic_Proposal} 
was performed in Hall C of the Thomas Jefferson National Accelerator Facility (Jefferson Lab) in 2020. 
The original goal was to determine the first diffractive minima of both electric and magnetic form factors of $^3$He by measuring the double spin asymmetry of $e-^3$He elastic scattering. 
Meanwhile, the experimental conditions provided an opportunity to also measure the double spin asymmetry of quasielastic scattering. 
A longitudinally polarized electron beam of 2.2 GeV energy was scattered off a 40-cm long polarized $^3$He target filled to about 12 amagats of density. The beam polarization $P_b$ was 
found to be $(85\pm 3)\%$, determined from Moller polarimetry. 

Scattered electrons were detected in the Super High Momentum Spectrometer (SHMS). The SHMS consists of five superconducting magnets \cite{HallC_Equipment} which follows the form of (DQQQD). The first dipole (D) is a horizontal bender (HB) used to bend scattered electrons by a small angle ($\approx$3 $^{\circ}$) and allows the spectrometer to reach a minimum scattering angle of $5.5^{\circ}$. The remaining quadruple (Q) elements are used for focusing of the scattered particles while the last dipole is used for momentum selection and deflects the particles into the SHMS detectors. 

The series of detectors in the SHMS provide position, timing, and particle identification of the scattered particles. 
The first detector is a pair of drift chambers that provide position and tracking information, 
which can be coupled with optics properties of the spectrometer to reconstruct the particles scattering angle, momentum, and the interaction position in the target. Following the drift chamber are two planes of hodoscopes, with each plane being comprised of vertical (Y) and horizontal (X) scintillitator paddles, and is used primarily as a trigger. A threshold gas Cerenkov detector and an electromagnetic calorimeter are used for particle identification. 

The target system included a gaseous polarized $^{3}$He target, along with an unpolarized reference cell, as well as additional solid targets used for calibration. The $^{3}$He target was polarized via spin exchange optical pumping \cite{Walker1997629,SinghPhysRevC_91}. To ascertain the target polarization, a two step process is used. First, the absolute polarization was determined through an Electron Paramagnetic Resonance (EPR) measurement, which served as a calibration point. Second, Adiabatic Fast Passage NMR measurements provide a relative polarization. NMR measurements can be taken more frequently during data collection, and when combined with the calibration constant from the EPR measurement, allows for a determination of the target polarization over time. The target polarization $P_t$ was found to vary 
between (35 - 45)\%, with $\Delta P_t/P_t=$ 3\%.  
Two tungsten collimators were placed adjacent to the target cell in order to reduce background events that scattered off the glass windows.

Data were collected at two primary kinematics, which provided asymmetries at four-momentum transfer squares $Q^{2}$ = 0.1 and $Q^{2}$ = 0.2 (GeV/c)$^{2}$. In order to better balance the proportion of elastic and quasielastic events, selected scintillator paddles in the SHMS were turned off, reducing the acceptance of the spectrometer.



Electrons were identified through particle identification (PID) cuts based on the gas Cerenkov and electromagnetic calorimeter detectors. A minimum response in the Cerenkov of 5 photoelectrons along with an $E/p$ great than 0.8, where $E$ is the particle energy measured by the calorimeter and $p$ is the reconstructed momentum, were required. Additionally, acceptance cuts 
were used to exclude regions of acceptance not well constrained by the existing optics matrix. Finally, a reconstructed position along the length of the target cell, $z_{targ}$ was optimized to suppress events which scattered from the glass windows of target cell. 

To form the asymmetry, yields were sorted based on the incident electron's helicity: 
\begin{align}
    Y^{+} = \frac{N^{+}}{Q^{+}LT^{+}} && Y^{-} = \frac{N^{-}}{Q^{-}LT^{-}},
\end{align}
where $N^{+}$ ($N^{-}$) represents the number of detected electrons in + (-) beam helicity state, $Q^{+}$ ($Q^{-}$) and $LT^{+}$ ($LT^{-}$) the corresponding total incident beam charge and the live time, respectively.  The asymmetry can then be formed as: 
\begin{equation}
    A_{raw} = \frac{Y^{+} - Y^{-} }{Y^{+} + Y^{-}}.
\end{equation}


To determine the acceptance and target cuts, a single-spectrometer Monte Carlo (SIMC) \cite{mcsingle} was adapted to describe the experimental setup. 
To ensure a satisfactory description of the scintillator paddle status, the simulated events were compared with actual data in both the focal plane and the reconstructed target quantities. The simulated events were weighted using elastic and quasielastic cross section models. The individual contributions from elastic and quasielastic scattering as well as the combined distribution were analyzed and showed a good agreement between data and simulation. With the observed good agreement, the momentum distribution was used to separate elastic from the quasielastic spectrum.



\begin{figure}
    \centering
    \includegraphics[scale=0.07]{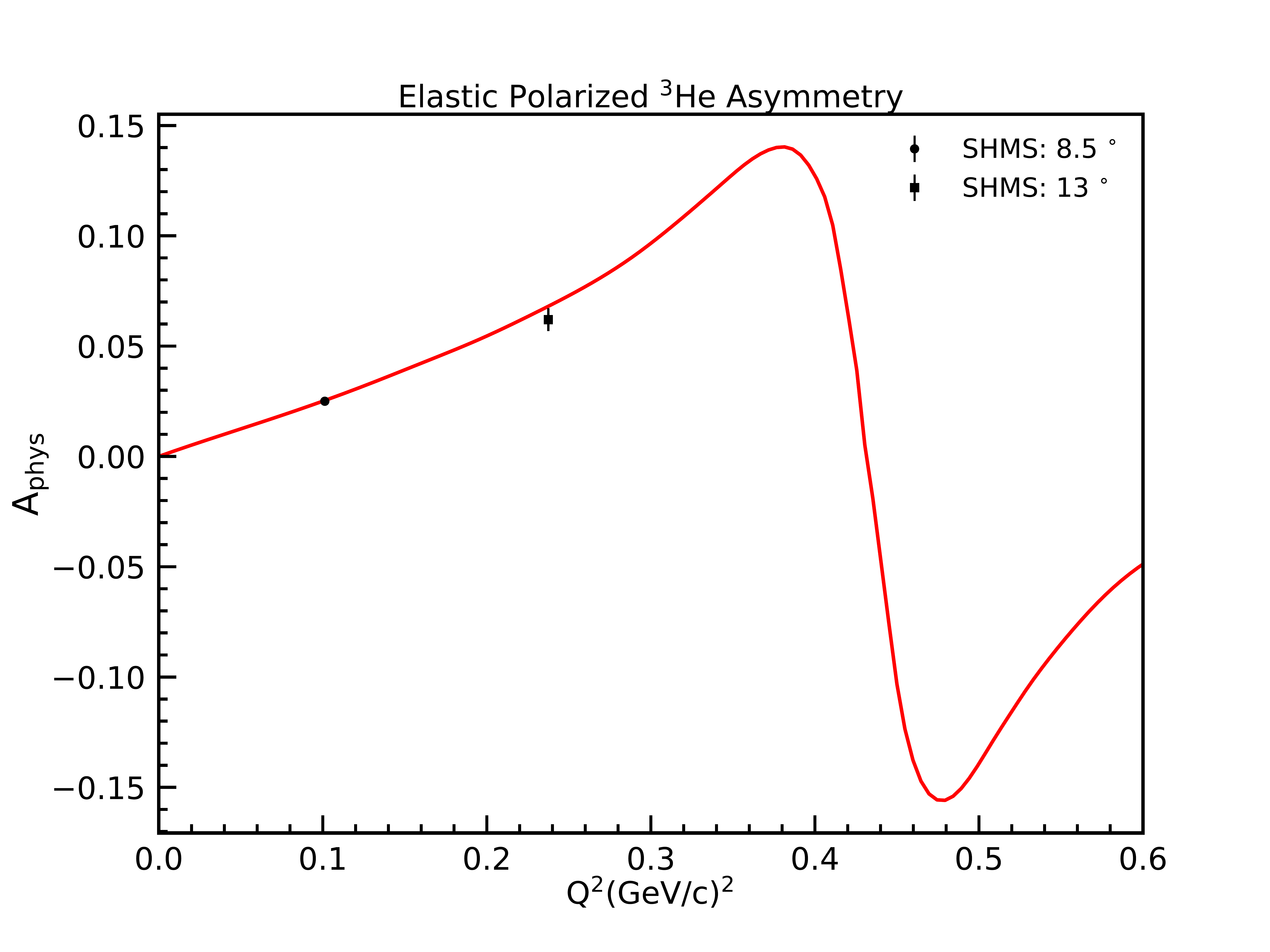}
    \caption{Measured elastic asymmetry for the $Q^{2}$ = 0.1 and $Q^{2}$ = 0.2 (GeV/$c$)$^2$(red solid circle), along with a prediction for the asymmetry (red curve). The error bar includes both statistical and systematic uncertainties}  
    \label{fig:Elastic_Asymmetry}
\end{figure}

 The elastic asymmetry extracted from data provided an independent check of the product of the beam and the target polarizations, $P_{b}P_{t}$. Illustrated in Fig.~\ref{fig:Elastic_Asymmetry} is the measured elastic asymmetry for the kinematic settings $Q^{2}$=0.1 and $Q^{2}$=0.2 (GeV/$c$)$^{2}$ along with a parameterization~\cite{BarcusThesis} for the elastic asymmetryutilizing form factors determined from a sum of Gaussians fits to world data~\cite{AMROUN1994596}. There is an overall good agreement between the measured asymmetries and predictions at low $Q^{2}$, where the elastic form factors are well constrained, indicating that the product $P_bP_t$ are reasonably well understood. Due to the limited beam time, however, the $Q^2$ coverage was not high enough to cover the diffractive minima. 

\begin{figure}
  \centering
    \includegraphics[width=.9\linewidth,height=180pt]{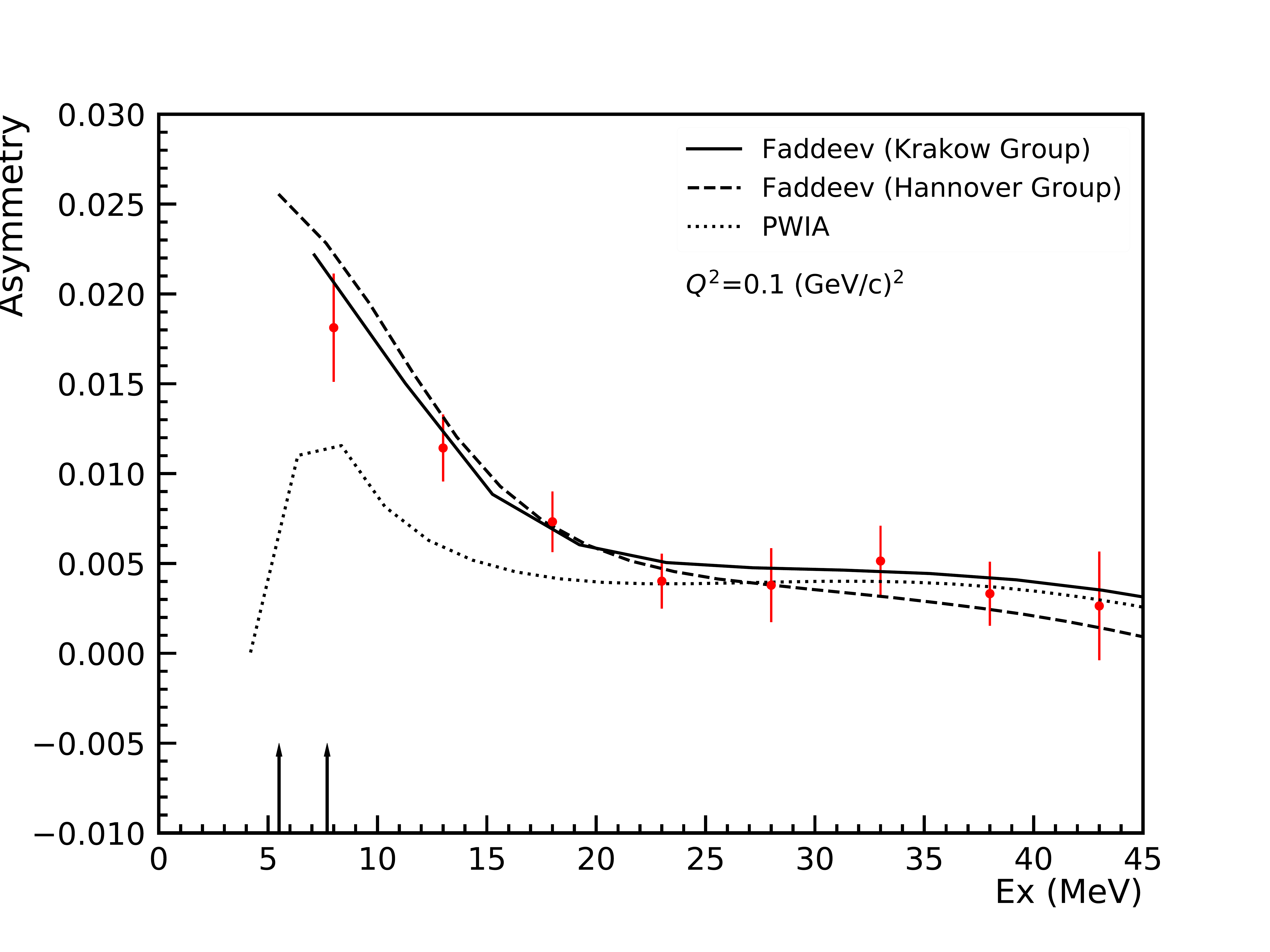} \\[\abovecaptionskip]
%
%
    \includegraphics[width=.9\linewidth,height=180pt]{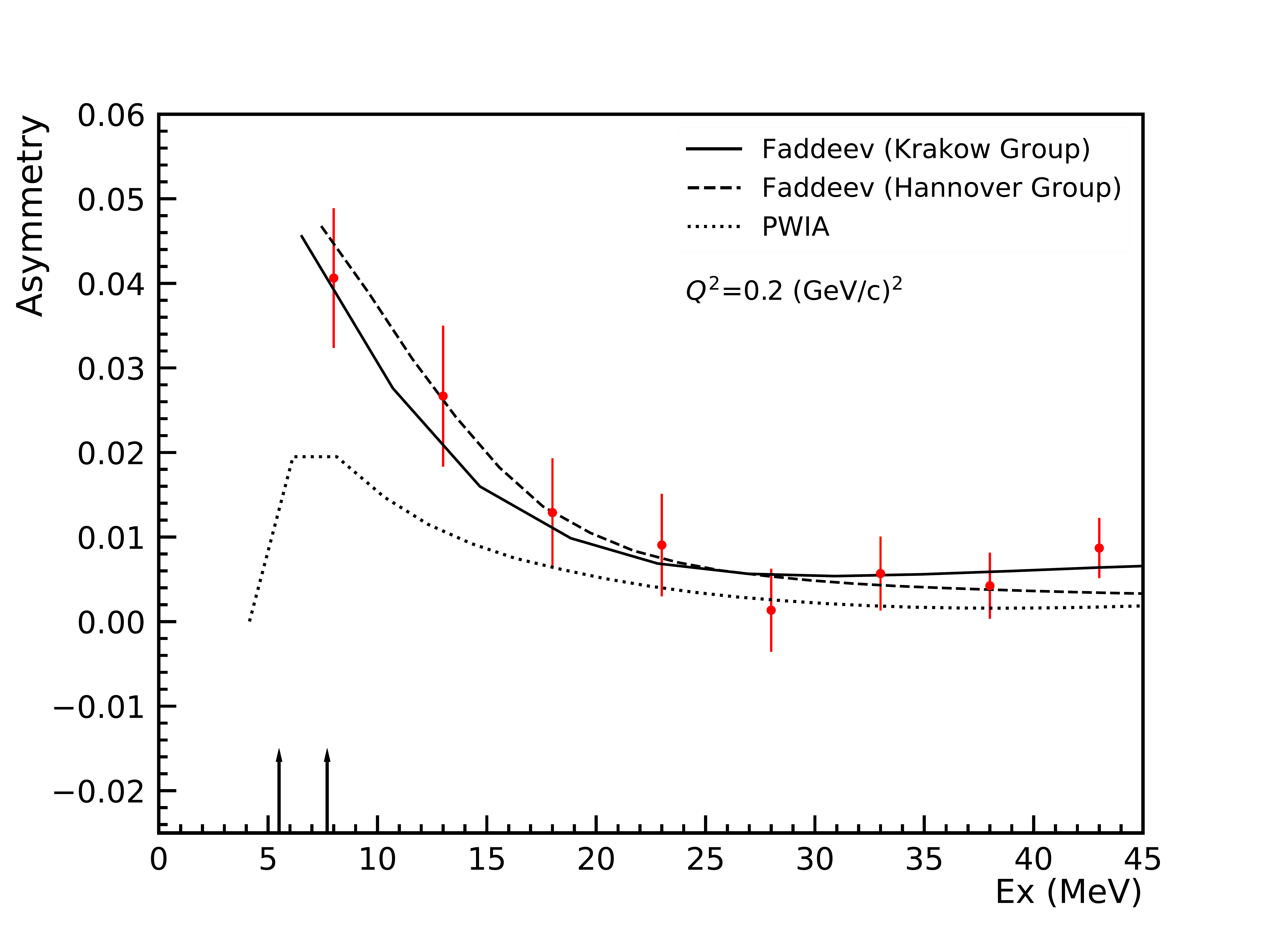} \\[\abovecaptionskip]
%
  \caption{Results on quasielastic asymmetry for $Q^{2}$=0.1 and $Q^{2}$=0.2~(GeV/$c$)$^2$, along with the theoretical predictions.}
  \label{fig:Phys_Asym}

\end{figure}

Combining the measured raw asymmetry with the beam and target polarizations, the QE physics asymmetry was obtained as: 
\begin{equation}
    A_{phys} = \frac{A_{raw}}{P_{b}P_{t}}\left( \frac{1}{1-d_{El}} \right) - A_{El} \left( \frac{d_{El}}{1 - d_{El}} \right),
\end{equation}
where $d_{El}$ and $A_{El}$ are the fractional contribution from and the asymmetry of elastic scattering. The extracted QE asymmetry was studied further as a function of the excitation energy of $^{3}$He, 
defined as
\begin{equation}
    E_x = \sqrt{M^{2} + 2M\omega - Q^{2}} - M,
\end{equation}
%
where $M$ is the mass of $^{3}$He and $\omega$ is the energy transfer. 
The two-body and three-body breakups correspond to excitation energies of 5.5 and 7.7 MeV, respectively. Results for the asymmetry are 
given in Tables \ref{table:kin_1} and \ref{table:kin_2}, along with both statistical and systematic uncertainties. The systematic uncertainties include contributions from elastic radiative tail subtraction, spectrometer acceptance, and beam \& target polarization ($P_{b}P_{t}$). The results are compared with three theoretical predictions in Fig.~\ref{fig:Phys_Asym}. 

\begin{table}[hbt!]
\begin{center}  
\caption{\label{Table:Q2=0.1} The measured asymmetry as a function of the excitation energy for $Q^{2}$=0.1 (GeV/c)$^{2}$. }
\begin{tabular}{p{2cm} wc{2.cm} wc{2.cm} wc{2.cm}}
\hline
\hline
\centering $E_x$ (MeV) & $A$  & $\delta_{stat}$ & $\delta_{syst}$ \\
\hline
\centering 8.0 & 0.0181 & 0.0027 & 0.0013\\ 
\centering 13.0 & 0.0114 & 0.0016 & 0.0008\\ 
 \centering 18.0 & 0.0073 & 0.0016 & 0.0005\\ 
 \centering 23.0 & 0.0040 & 0.0015 & 0.0003\\ 
 \centering 28.0 & 0.0038 & 0.0020 & 0.0003\\ 
 \centering 33.0 & 0.0051 & 0.0019 & 0.0004\\ 
\centering 38.0 & 0.0033 & 0.0018 & 0.0002\\ 
\centering 43.0 & 0.0026 & 0.0030 & 0.0002\\ 
\hline
\label{table:kin_1}
\end{tabular}
\end{center}
\end{table}

\begin{table}[hbt!]
\begin{center}  
\caption{\label{Table:Q2=0.1} The measured asymmetry as a function of the excitation energy for $Q^{2}$=0.2 (GeV/c)$^{2}$. }
\setlength\doublerulesep{1pt}
\begin{tabular}{wc{2cm} wc{2.cm} wc{2.cm} wc{2.cm}}
\hline
\hline
\rule{0pt}{10pt}
\centering $E_x$ (MeV) & $A$  & $\delta_{stat}$ & $\delta_{syst}$ \rule[-5pt]{0pt}{0pt} \\
\hline
\rule{0pt}{10pt}
\centering 8.0 & 0.0404 & 0.0023 & 0.0029\\ 
\centering 13.0 & 0.0266 & 0.0025 & 0.0019\\ 
 \centering 18.0 & 0.0128 & 0.0019 & 0.0009\\ 
 \centering 23.0 & 0.0090 & 0.0018 & 0.0006\\ 
 \centering 28.0 & 0.0014 & 0.0015 & 0.0001\\ 
 \centering 33.0 & 0.0057 & 0.0013 & 0.0004\\ 
\centering 38.0 & 0.0042 & 0.0012 & 0.0003\\ 
\centering 43.0 & 0.0086 & 0.0010 & 0.0006\\ 
\hline
\label{table:kin_2}
\end{tabular}
\end{center}
\end{table}

Of the three models, one is the relativistic plane-wave impulse approximation \cite{PWIA1} that evaluates the spectral function utilizing the AV18 nucleon-nucleon potential \cite{AV18}, along with a Coulomb potential and takes the H{\"o}hler parameterization \cite{HOHLER1976505} for the form factors. 
PWIA is generally not expected to be reliable in the three-nucleon system at low energies where rescattering and FSI are important. 
The two remaining models  \cite{Golak1995,Deltuva2004} are based on the nonrelativistic Faddeev theory for three-particle scattering and include both FSI and MEC. They differ in the choice of the nuclear interaction, i.e., the AV18 potential with the Urbana IX 3NF was used in the calculations by Golak {\it et al.} \cite{GOLAK200589} while the CD Bonn + $\Delta$ potential including an explicit $\Delta$-isobar excitation and the proton-proton Coulomb force was used in the calculations by Deltuva {\it et al.} \cite{Deltuva2004,DELTUVA2022137552}. Both Faddeev calculations utilized electromagnetic form factors from Hammer {\it et al.} \cite{Hammer2004}. 
As can be seen in Figure \ref{fig:Phys_Asym}, both nonrelativistic Faddeev calculations  
have an overall good agreement with data at both kinematics, while PWIA fails at lower excitation energies. This finding is consistent with a previous study at low $Q^{2}$, where the Faddeev calculations could explain the experimental data \cite{Feng_Xiong} considerably better than PWIA.

We report here a recent, precision measurement of the double-spin asymmetry in electron quasielastic scattering off a $^3$He target. The asymmetries are presented as functions of the excitation energy and are compared with theoretical calculations using PWIA and Fadeev equations with two different nucleon-nucleon potential inputs. Both Fadeev-based calculations agree with the new data fairly well, though the PWIA agrees with data only at high excitation energies. 
We expect that the new data set, taken at relatively low energy and momentum transfer, will be confronted with incoming predictions based on 2N and 3N potentials as well as electromagnetic current operators derived consistently within chiral effective field theory. 





\begin{acknowledgments}
We thank the Accelerator Division and Hall C technical staff for their support. 
This material is based upon work supported in part by the U.S. Department of Energy, Office of Science, Office of Nuclear Physics under contract numbers DE-AC05-06OR23177 (JLab), DE-FG02-99ER41101 (Univ. of Kentucky), 
and 
DE–SC0014434 (U. of Virginia). 
%
%
The work of J. Golak, R. Skibi\'nski, and H. Witała (Krak\'ow group) was supported by the ExcellenceInitiative - Research University Program at the Jagiellonian University in Krak\'ow. The numerical calculations of the Krak\'ow group were partly performed on the supercomputers of the JSC, J\"ulich, Germany.
%
\end{acknowledgments}


\appendix


\bibliography{apssamp.bib}

\providecommand{\noopsort}[1]{}\providecommand{\singleletter}[1]{#1}%
\begin{thebibliography}{22}%
\makeatletter
\providecommand \@ifxundefined [1]{%
 \@ifx{#1\undefined}
}%
\providecommand \@ifnum [1]{%
 \ifnum #1\expandafter \@firstoftwo
 \else \expandafter \@secondoftwo
 \fi
}%
\providecommand \@ifx [1]{%
 \ifx #1\expandafter \@firstoftwo
 \else \expandafter \@secondoftwo
 \fi
}%
\providecommand \natexlab [1]{#1}%
\providecommand \enquote  [1]{``#1''}%
\providecommand \bibnamefont  [1]{#1}%
\providecommand \bibfnamefont [1]{#1}%
\providecommand \citenamefont [1]{#1}%
\providecommand \href@noop [0]{\@secondoftwo}%
\providecommand \href [0]{\begingroup \@sanitize@url \@href}%
\providecommand \@href[1]{\@@startlink{#1}\@@href}%
\providecommand \@@href[1]{\endgroup#1\@@endlink}%
\providecommand \@sanitize@url [0]{\catcode `\\12\catcode `\$12\catcode `\&12\catcode `\#12\catcode `\^12\catcode `\_12\catcode `\%12\relax}%
\providecommand \@@startlink[1]{}%
\providecommand \@@endlink[0]{}%
\providecommand \url  [0]{\begingroup\@sanitize@url \@url }%
\providecommand \@url [1]{\endgroup\@href {#1}{\urlprefix }}%
\providecommand \urlprefix  [0]{URL }%
\providecommand \Eprint [0]{\href }%
\providecommand \doibase [0]{https://doi.org/}%
\providecommand \selectlanguage [0]{\@gobble}%
\providecommand \bibinfo  [0]{\@secondoftwo}%
\providecommand \bibfield  [0]{\@secondoftwo}%
\providecommand \translation [1]{[#1]}%
\providecommand \BibitemOpen [0]{}%
\providecommand \bibitemStop [0]{}%
\providecommand \bibitemNoStop [0]{.\EOS\space}%
\providecommand \EOS [0]{\spacefactor3000\relax}%
\providecommand \BibitemShut  [1]{\csname bibitem#1\endcsname}%
\let\auto@bib@innerbib\@empty
\bibitem [{\citenamefont {Mihovilovič}\ \emph {et~al.}(2019)\citenamefont {Mihovilovič} \emph {et~al.}}]{Miha_2014}%
  \BibitemOpen
  \bibfield  {author} {\bibinfo {author} {\bibfnamefont {M.}~\bibnamefont {Mihovilovič}} \emph {et~al.},\ }\href {https://doi.org/https://doi.org/10.1016/j.physletb.2018.10.063} {\bibfield  {journal} {\bibinfo  {journal} {Physics Letters B}\ }\textbf {\bibinfo {volume} {788}},\ \bibinfo {pages} {117} (\bibinfo {year} {2019})}\BibitemShut {NoStop}%
\bibitem [{\citenamefont {Ciofi~degli Atti}\ \emph {et~al.}(1995)\citenamefont {Ciofi~degli Atti}, \citenamefont {Pace},\ and\ \citenamefont {Salm\`e}}]{PWIA1}%
  \BibitemOpen
  \bibfield  {author} {\bibinfo {author} {\bibfnamefont {C.}~\bibnamefont {Ciofi~degli Atti}}, \bibinfo {author} {\bibfnamefont {E.}~\bibnamefont {Pace}},\ and\ \bibinfo {author} {\bibfnamefont {G.}~\bibnamefont {Salm\`e}},\ }\bibfield  {journal} {\bibinfo  {journal} {Phys. Rev. C}\ }\textbf {\bibinfo {volume} {1}},\ \href {https://doi.org/10.1103/PhysRevC.51.1108} {10.1103/PhysRevC.51.1108} (\bibinfo {year} {1995})\BibitemShut {NoStop}%
\bibitem [{\citenamefont {Kievsky}\ \emph {et~al.}(1997)\citenamefont {Kievsky}, \citenamefont {Pace}, \citenamefont {Salm\`e},\ and\ \citenamefont {Viviani}}]{PWIA2}%
  \BibitemOpen
  \bibfield  {author} {\bibinfo {author} {\bibfnamefont {A.}~\bibnamefont {Kievsky}}, \bibinfo {author} {\bibfnamefont {E.}~\bibnamefont {Pace}}, \bibinfo {author} {\bibfnamefont {G.}~\bibnamefont {Salm\`e}},\ and\ \bibinfo {author} {\bibfnamefont {M.}~\bibnamefont {Viviani}},\ }\href {https://doi.org/10.1103/PhysRevC.56.64} {\bibfield  {journal} {\bibinfo  {journal} {Phys. Rev. C}\ }\textbf {\bibinfo {volume} {56}},\ \bibinfo {pages} {64} (\bibinfo {year} {1997})}\BibitemShut {NoStop}%
\bibitem [{\citenamefont {Pace}\ \emph {et~al.}(2001)\citenamefont {Pace}, \citenamefont {Salm\`e}, \citenamefont {Scopetta},\ and\ \citenamefont {Kievsky}}]{PhysRevC.64.055203}%
  \BibitemOpen
  \bibfield  {author} {\bibinfo {author} {\bibfnamefont {E.}~\bibnamefont {Pace}}, \bibinfo {author} {\bibfnamefont {G.}~\bibnamefont {Salm\`e}}, \bibinfo {author} {\bibfnamefont {S.}~\bibnamefont {Scopetta}},\ and\ \bibinfo {author} {\bibfnamefont {A.}~\bibnamefont {Kievsky}},\ }\href {https://doi.org/10.1103/PhysRevC.64.055203} {\bibfield  {journal} {\bibinfo  {journal} {Phys. Rev. C}\ }\textbf {\bibinfo {volume} {64}},\ \bibinfo {pages} {055203} (\bibinfo {year} {2001})}\BibitemShut {NoStop}%
\bibitem [{\citenamefont {Golak}\ \emph {et~al.}(1995)\citenamefont {Golak}, \citenamefont {Wital}, \citenamefont {Kamada}, \citenamefont {H\"uber}, \citenamefont {Ishikawa},\ and\ \citenamefont {Gl\"ockle}}]{Golak1995}%
  \BibitemOpen
  \bibfield  {author} {\bibinfo {author} {\bibfnamefont {J.}~\bibnamefont {Golak}}, \bibinfo {author} {\bibfnamefont {H.}~\bibnamefont {Wital}}, \bibinfo {author} {\bibfnamefont {H.}~\bibnamefont {Kamada}}, \bibinfo {author} {\bibfnamefont {D.}~\bibnamefont {H\"uber}}, \bibinfo {author} {\bibfnamefont {S.}~\bibnamefont {Ishikawa}},\ and\ \bibinfo {author} {\bibfnamefont {W.}~\bibnamefont {Gl\"ockle}},\ }\href {https://doi.org/10.1103/PhysRevC.52.1216} {\bibfield  {journal} {\bibinfo  {journal} {Phys. Rev. C}\ }\textbf {\bibinfo {volume} {52}},\ \bibinfo {pages} {1216} (\bibinfo {year} {1995})}\BibitemShut {NoStop}%
\bibitem [{\citenamefont {Golak}\ \emph {et~al.}(2005)\citenamefont {Golak}, \citenamefont {Skibiński}, \citenamefont {Witała}, \citenamefont {Glöckle}, \citenamefont {Nogga},\ and\ \citenamefont {Kamada}}]{GOLAK200589}%
  \BibitemOpen
  \bibfield  {author} {\bibinfo {author} {\bibfnamefont {J.}~\bibnamefont {Golak}}, \bibinfo {author} {\bibfnamefont {R.}~\bibnamefont {Skibiński}}, \bibinfo {author} {\bibfnamefont {H.}~\bibnamefont {Witała}}, \bibinfo {author} {\bibfnamefont {W.}~\bibnamefont {Glöckle}}, \bibinfo {author} {\bibfnamefont {A.}~\bibnamefont {Nogga}},\ and\ \bibinfo {author} {\bibfnamefont {H.}~\bibnamefont {Kamada}},\ }\href {https://doi.org/https://doi.org/10.1016/j.physrep.2005.04.005} {\bibfield  {journal} {\bibinfo  {journal} {Physics Reports}\ }\textbf {\bibinfo {volume} {415}},\ \bibinfo {pages} {89} (\bibinfo {year} {2005})}\BibitemShut {NoStop}%
\bibitem [{\citenamefont {Deltuva}\ \emph {et~al.}(2004)\citenamefont {Deltuva}, \citenamefont {Yuan}, \citenamefont {Adam},\ and\ \citenamefont {Sauer}}]{Deltuva2004}%
  \BibitemOpen
  \bibfield  {author} {\bibinfo {author} {\bibfnamefont {A.}~\bibnamefont {Deltuva}}, \bibinfo {author} {\bibfnamefont {L.~P.}\ \bibnamefont {Yuan}}, \bibinfo {author} {\bibfnamefont {J.}~\bibnamefont {Adam}},\ and\ \bibinfo {author} {\bibfnamefont {P.~U.}\ \bibnamefont {Sauer}},\ }\href {https://doi.org/10.1103/PhysRevC.70.034004} {\bibfield  {journal} {\bibinfo  {journal} {Phys. Rev. C}\ }\textbf {\bibinfo {volume} {70}},\ \bibinfo {pages} {034004} (\bibinfo {year} {2004})}\BibitemShut {NoStop}%
\bibitem [{\citenamefont {Deltuva}(2022)}]{DELTUVA2022137552}%
  \BibitemOpen
  \bibfield  {author} {\bibinfo {author} {\bibfnamefont {A.}~\bibnamefont {Deltuva}},\ }\href {https://doi.org/https://doi.org/10.1016/j.physletb.2022.137552} {\bibfield  {journal} {\bibinfo  {journal} {Physics Letters B}\ }\textbf {\bibinfo {volume} {835}},\ \bibinfo {pages} {137552} (\bibinfo {year} {2022})}\BibitemShut {NoStop}%
\bibitem [{\citenamefont {Wita\l{}a}\ \emph {et~al.}(2023)\citenamefont {Wita\l{}a}, \citenamefont {Golak},\ and\ \citenamefont {Skibi\'nski}}]{witała2023inclusion}%
  \BibitemOpen
  \bibfield  {author} {\bibinfo {author} {\bibfnamefont {H.}~\bibnamefont {Wita\l{}a}}, \bibinfo {author} {\bibfnamefont {J.}~\bibnamefont {Golak}},\ and\ \bibinfo {author} {\bibfnamefont {R.}~\bibnamefont {Skibi\'nski}},\ }\href@noop {} {} (\bibinfo {year} {2023}),\ \Eprint {https://arxiv.org/abs/2310.03433} {arXiv:2310.03433 [nucl-th]} \BibitemShut {NoStop}%
\bibitem [{\citenamefont {Retzlaff}\ \emph {et~al.}(1994)\citenamefont {Retzlaff} \emph {et~al.}}]{PhysRevC.49.1263}%
  \BibitemOpen
  \bibfield  {author} {\bibinfo {author} {\bibfnamefont {G.~A.}\ \bibnamefont {Retzlaff}} \emph {et~al.},\ }\href {https://doi.org/10.1103/PhysRevC.49.1263} {\bibfield  {journal} {\bibinfo  {journal} {Phys. Rev. C}\ }\textbf {\bibinfo {volume} {49}},\ \bibinfo {pages} {1263} (\bibinfo {year} {1994})}\BibitemShut {NoStop}%
\bibitem [{\citenamefont {Xiong}\ \emph {et~al.}(2001)\citenamefont {Xiong} \emph {et~al.}}]{Feng_Xiong}%
  \BibitemOpen
  \bibfield  {author} {\bibinfo {author} {\bibfnamefont {F.}~\bibnamefont {Xiong}} \emph {et~al.},\ }\href {https://doi.org/10.1103/PhysRevLett.87.242501} {\bibfield  {journal} {\bibinfo  {journal} {Phys. Rev. Lett.}\ }\textbf {\bibinfo {volume} {87}},\ \bibinfo {pages} {242501} (\bibinfo {year} {2001})}\BibitemShut {NoStop}%
\bibitem [{\citenamefont {Donnelly}\ and\ \citenamefont {Raskin}(1986)}]{Donnelly_pol}%
  \BibitemOpen
  \bibfield  {author} {\bibinfo {author} {\bibfnamefont {T.~W.}\ \bibnamefont {Donnelly}}\ and\ \bibinfo {author} {\bibfnamefont {A.~S.}\ \bibnamefont {Raskin}},\ }\href@noop {} {\bibfield  {journal} {\bibinfo  {journal} {Annals of Physics}\ } (\bibinfo {year} {1986})}\BibitemShut {NoStop}%
\bibitem [{\citenamefont {S.~Barcus}\ \emph {et~al.}(2020)\citenamefont {S.~Barcus} \emph {et~al.}}]{Elastic_Proposal}%
  \BibitemOpen
  \bibfield  {author} {\bibinfo {author} {\bibfnamefont {S.~L.}\ \bibnamefont {S.~Barcus}, \bibfnamefont {D.~Higinbotham}} \emph {et~al.},\ }\href {https://hallcweb.jlab.org/wiki/images/b/bc/PAC_Proposal_3He_Polarization_Observables.pdf} {\bibinfo {title} {Measurement of $^3\mathrm{He}$ elastic electromagnetic form factor diffractive minima using polarization observables}} (\bibinfo {year} {2020})\BibitemShut {NoStop}%
\bibitem [{\citenamefont {Wood}\ \emph {et~al.}(2019)\citenamefont {Wood} \emph {et~al.}}]{HallC_Equipment}%
  \BibitemOpen
  \bibfield  {author} {\bibinfo {author} {\bibfnamefont {S.}~\bibnamefont {Wood}} \emph {et~al.},\ }\href {https://hallcweb.jlab.org/safety-docs/current/Standard-Equipment-Manual.pdf} {\bibinfo {title} {{Jefferson Lab Hall C Standard Equipment Manual}}} (\bibinfo {year} {2019})\BibitemShut {NoStop}%
\bibitem [{\citenamefont {Walker}\ and\ \citenamefont {Happer}(1997)}]{Walker1997629}%
  \BibitemOpen
  \bibfield  {author} {\bibinfo {author} {\bibfnamefont {T.~G.}\ \bibnamefont {Walker}}\ and\ \bibinfo {author} {\bibfnamefont {W.}~\bibnamefont {Happer}},\ }\href {https://doi.org/10.1103/revmodphys.69.629} {\bibfield  {journal} {\bibinfo  {journal} {Reviews of Modern Physics}\ }\textbf {\bibinfo {volume} {69}},\ \bibinfo {pages} {629 – 642} (\bibinfo {year} {1997})}\BibitemShut {NoStop}%
\bibitem [{\citenamefont {Singh}\ \emph {et~al.}(2015)\citenamefont {Singh}, \citenamefont {Dolph}, \citenamefont {Tobias}, \citenamefont {Averett}, \citenamefont {Kelleher}, \citenamefont {Mooney}, \citenamefont {Nelyubin}, \citenamefont {Wang}, \citenamefont {Zheng},\ and\ \citenamefont {Cates}}]{SinghPhysRevC_91}%
  \BibitemOpen
  \bibfield  {author} {\bibinfo {author} {\bibfnamefont {J.~T.}\ \bibnamefont {Singh}}, \bibinfo {author} {\bibfnamefont {P.~A.~M.}\ \bibnamefont {Dolph}}, \bibinfo {author} {\bibfnamefont {W.~A.}\ \bibnamefont {Tobias}}, \bibinfo {author} {\bibfnamefont {T.~D.}\ \bibnamefont {Averett}}, \bibinfo {author} {\bibfnamefont {A.}~\bibnamefont {Kelleher}}, \bibinfo {author} {\bibfnamefont {K.~E.}\ \bibnamefont {Mooney}}, \bibinfo {author} {\bibfnamefont {V.~V.}\ \bibnamefont {Nelyubin}}, \bibinfo {author} {\bibfnamefont {Y.}~\bibnamefont {Wang}}, \bibinfo {author} {\bibfnamefont {Y.}~\bibnamefont {Zheng}},\ and\ \bibinfo {author} {\bibfnamefont {G.~D.}\ \bibnamefont {Cates}},\ }\href {https://doi.org/10.1103/PhysRevC.91.055205} {\bibfield  {journal} {\bibinfo  {journal} {Phys. Rev. C}\ }\textbf {\bibinfo {volume} {91}},\ \bibinfo {pages} {055205} (\bibinfo {year} {2015})}\BibitemShut {NoStop}%
\bibitem [{mcs(2021)}]{mcsingle}%
  \BibitemOpen
  \href@noop {} {\bibinfo {title} {{Hall C SHMS Single Arm Monte Carlo}}},\ \bibinfo {howpublished} {\url{https://github.com/JeffersonLab/mc_shms_single}} (\bibinfo {year} {2021})\BibitemShut {NoStop}%
\bibitem [{\citenamefont {Barcus}(2019)}]{BarcusThesis}%
  \BibitemOpen
  \bibfield  {author} {\bibinfo {author} {\bibfnamefont {S.}~\bibnamefont {Barcus}},\ }\href@noop {} {\bibinfo {type} {{Ph.D.} thesis}},\ \bibinfo  {school} {College of William \& Mary} (\bibinfo {year} {2019})\BibitemShut {NoStop}%
\bibitem [{\citenamefont {Amroun}\ \emph {et~al.}(1994)\citenamefont {Amroun} \emph {et~al.}}]{AMROUN1994596}%
  \BibitemOpen
  \bibfield  {author} {\bibinfo {author} {\bibfnamefont {A.}~\bibnamefont {Amroun}} \emph {et~al.},\ }\href {https://doi.org/https://doi.org/10.1016/0375-9474(94)90925-3} {\bibfield  {journal} {\bibinfo  {journal} {Nuclear Physics A}\ }\textbf {\bibinfo {volume} {579}},\ \bibinfo {pages} {596} (\bibinfo {year} {1994})}\BibitemShut {NoStop}%
\bibitem [{\citenamefont {Wiringa}\ \emph {et~al.}(1995)\citenamefont {Wiringa}, \citenamefont {Stoks},\ and\ \citenamefont {Schiavilla}}]{AV18}%
  \BibitemOpen
  \bibfield  {author} {\bibinfo {author} {\bibfnamefont {R.~B.}\ \bibnamefont {Wiringa}}, \bibinfo {author} {\bibfnamefont {V.~G.~J.}\ \bibnamefont {Stoks}},\ and\ \bibinfo {author} {\bibfnamefont {R.}~\bibnamefont {Schiavilla}},\ }\href {https://doi.org/10.1103/PhysRevC.51.38} {\bibfield  {journal} {\bibinfo  {journal} {Phys. Rev. C}\ }\textbf {\bibinfo {volume} {51}},\ \bibinfo {pages} {38} (\bibinfo {year} {1995})}\BibitemShut {NoStop}%
\bibitem [{\citenamefont {Höhler}\ \emph {et~al.}(1976)\citenamefont {Höhler}, \citenamefont {Pietarinen}, \citenamefont {Sabba-Stefanescu}, \citenamefont {Borkowski}, \citenamefont {Simon}, \citenamefont {Walther},\ and\ \citenamefont {Wendling}}]{HOHLER1976505}%
  \BibitemOpen
  \bibfield  {author} {\bibinfo {author} {\bibfnamefont {G.}~\bibnamefont {Höhler}}, \bibinfo {author} {\bibfnamefont {E.}~\bibnamefont {Pietarinen}}, \bibinfo {author} {\bibfnamefont {I.}~\bibnamefont {Sabba-Stefanescu}}, \bibinfo {author} {\bibfnamefont {F.}~\bibnamefont {Borkowski}}, \bibinfo {author} {\bibfnamefont {G.}~\bibnamefont {Simon}}, \bibinfo {author} {\bibfnamefont {V.}~\bibnamefont {Walther}},\ and\ \bibinfo {author} {\bibfnamefont {R.}~\bibnamefont {Wendling}},\ }\href {https://doi.org/https://doi.org/10.1016/0550-3213(76)90449-1} {\bibfield  {journal} {\bibinfo  {journal} {Nuclear Physics B}\ }\textbf {\bibinfo {volume} {114}},\ \bibinfo {pages} {505} (\bibinfo {year} {1976})}\BibitemShut {NoStop}%
\bibitem [{\citenamefont {Hammer}\ and\ \citenamefont {Meißner}(2004)}]{Hammer2004}%
  \BibitemOpen
  \bibfield  {author} {\bibinfo {author} {\bibfnamefont {H.-W.}\ \bibnamefont {Hammer}}\ and\ \bibinfo {author} {\bibfnamefont {U.-G.}\ \bibnamefont {Meißner}},\ }\href {https://doi.org/10.1140/epja/i2003-10223-y} {\bibfield  {journal} {\bibinfo  {journal} {Eur. Phys. J. A 20, 469–473 (2004)}\ }\textbf {\bibinfo {volume} {20}},\ \bibinfo {pages} {469} (\bibinfo {year} {2004})}\BibitemShut {NoStop}%
\end{thebibliography}%
\end{document}